\documentclass[aps,physrev,preprint,groupedaddress]{revtex4-2}
\usepackage{graphicx}
\usepackage{siunitx}
\usepackage{amsmath}
\usepackage{bm}
\usepackage{tikz}
\usepackage{tabularray}
\usepackage{float}

\newcommand{\vb}[1]{\mathbf{#1}}
\newcommand{\Ftot}{\mathbf{F}_{\rm tot}}
\newcommand{\Fx}{{\rm F_x}}
\newcommand{\Fy}{{\rm F_y}}
\newcommand{\Fz}{{\rm F_z}}
\newcommand{\wx}{{\rm w_x}}
\newcommand{\wy}{{\rm w_y}}
\newcommand{\Px}{{\rm \Lambda_x}}
\newcommand{\Py}{{\rm \Lambda_y}}

\begin{document}

%\preprint{}

\title{An exploration of lateral optical forces from a triangular periodic motif}


\author{Bo Gao}
%\thanks{}
\affiliation{H.H. Wills Physics Laboratory, University of Bristol, BS8~1TL, Bristol, UK}

\author{Henkjan Gersen}
%\thanks{}
\affiliation{iLof -- Intelligent Lab on Fibre, Porto 4300-240, Portugal}
\affiliation{H.H. Wills Physics Laboratory, University of Bristol, BS8~1TL, Bristol, UK}
% \altaffiliation[also at]{H.H. Wills Physics Laboratory, University of Bristol, BS8~1TL, Bristol, UK}

\author{Simon Hanna}
\email[Contact author: ]{s.hanna@bristol.ac.uk}
%\thanks{}
\affiliation{H.H. Wills Physics Laboratory, University of Bristol, BS8~1TL, Bristol, UK}


\date{\today}

\begin{abstract}
This computational study investigates lateral optical forces in asymmetric dielectric nanostructures, focusing on their connection to resonant light--matter interactions. 
We examine isosceles triangular motifs that exhibit two distinct types of optical force response under plane wave illumination. 
Through parameter-space analysis, we identify stable zones where optical forces remain consistent and switching bands where forces change abruptly as parameters are altered. 
The observed force spectra show characteristic asymmetric lineshapes, suggesting Fano-resonance behavior. 
Eigenfrequency analysis confirms these effects arise from interference between discrete eigenmodes and continuum propagation states, with the eigenmode Q-factors correlating with transition sharpness. 
These findings provide insights into how structural geometry influences optical forces through resonant effects, offering guidance for designing optically-driven systems where controlled optical force responses are desired.
\end{abstract}

%\keywords{}

\maketitle

\section{Introduction}

Lateral optical forces (LOFs) arising from the breaking of mirror symmetry in optical systems have emerged as a promising mechanism for enabling next-generation technologies, including precise sensors \cite{OptProbe}, nanoscale actuators \cite{Nanomotor}, and micro-robotics \cite{MetaV}.
Additionally, LOF enables advanced optical manipulations such as particle sorting \cite{OptSorting}, optical lifting \cite{Lift1, Lift2}, and self-stabilized levitation \cite{LightSail}, highlighting the fundamental role of symmetry-breaking in light--matter interactions.

Symmetry breaking can occur in the light field (e.g., through polarization, oblique incidence, or structured beams), the material system (e.g., via chirality, anisotropy, or geometric asymmetry), or a combination of both.
The scope of this work focuses on LOFs generated by geometric asymmetry in the spatial distribution of simple isotropic materials---a scenario commonly encountered in studies of metamaterials, nanophotonics, and optical manipulation systems.
Recent examples include particle pairs with mismatched refractive indices \cite{MixRI} or sizes \cite{MixShape, MetaV}, as well as liquid crystals with space-variant geometric phase elements \cite{MacroLC}.

\begin{figure}
    % \centering
    \includegraphics[width=0.8\linewidth]{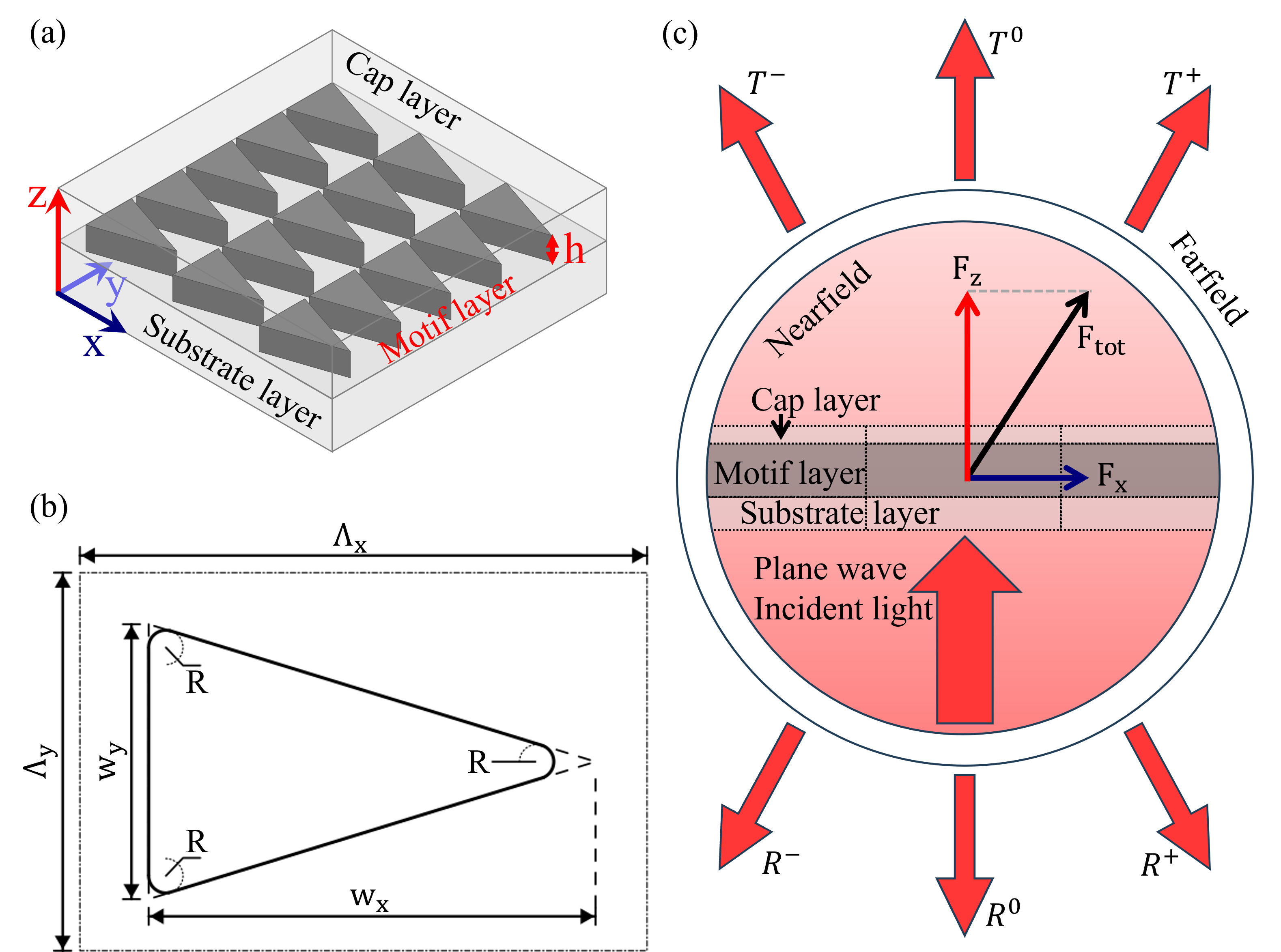}
    \caption{
        (a) A schematic view of the triangle model and coordinate systems. 
        Periodic boundary conditions are applied along the x- and y-directions.
        The structure consists of three layers: the substrate layer, the middle motif layer, and the capping layer.
        The isosceles triangles in the motif layer are oriented to be symmetric in the y-direction. 
        (b) A top-view presentation of the triangle shape in the motif layer. 
        Definitions of major shape parameters are denoted in the figure. 
        (c) A schematic view of the optical system in the X-Z plane. 
        The incident light is a plane wave linearly polarized in the x-direction, propagating along the z-direction.
        The diffraction efficiencies (DEs) in both reflection and transmission sides, as well as the total optical force $\Ftot$ and its x- and z-components, are presented.
    \label{fig:scheme}
    }
\end{figure}

Building on previous investigations of LOFs from asymmetric planar structures \cite{Triangle-1, Triangle-2}, we examine LOFs generated from breaking symmetry using a periodic array of all-dielectric isosceles triangular motifs as a model system (see Figure~\ref{fig:scheme}a).
The isosceles triangle shape is chosen because it preserves a single mirror symmetry, while exhibiting no symmetry in the orthogonal direction.
The lack of mirror symmetry thereby provides a clear indication of the directional orientation of the resulting LOFs.
Furthermore, most conventional asymmetric planar structures offer a high degree of freedom in spatial design.
They typically rely on particle pairs, multiple elements, or variations of highly symmetric shapes such as squares or circles.
In contrast, the triangular motif is built from a single asymmetric particle, and its planar triangular shape can be described by only two parameters.
Although an additional parameter, the motif thickness, plays an important role in the system, it is intentionally held fixed for most of the study and treated as an external factor not considered within the design space.

In an effort to probe the maximum achievable LOFs in the triangular motif system, we employed Bayesian optimization---a method renowned for efficiently finding near-global optima in high-dimensional, expensive-to-evaluate systems---to perform geometry optimization.
The results of Bayesian optimization revealed two counterintuitive effects: abrupt reversals in force direction from minimal geometric modifications, and near-identical forces produced by visually distinct geometries.
To understand this, we systematically investigated the landscape of LOFs in terms of geometric parameters, observing a ramp-and-step pattern appearing in several regions of the landscape.
From a closer examination of the optical force behavior, while varying the incident wavelength and physical parameters, we identified the ramp-and-step pattern to be characteristic of Fano resonances, providing novel insights into potential applications and fabrication strategies.
By analyzing the shift in eigenfrequencies as a response to changing geometric parameters, we verified the association with Fano resonances, and showed that it is possible to engineer the eigenmodes of the system to manipulate the optical force for specific applications for example in optical transducers or sensors.
% Section~\ref{sec:univ} examines the versatility of the ramp-and-step pattern and its relationship with Fano resonances by varying an additional factor unconsidered in previous sections, which implies a more fundamental nature of such a discovery.  

%----------------------------------------------------------------------------------------
% \section{Methodology \label{sec:method}}
\section{Model System \& LOF Calculation \label{sec:method}}

To support our aims to examine the LOF generated from the geometric asymmetry of the periodic triangle motifs, a 3D model system suitable for fabrication and a computationally efficient method to simulate LOFs for each given structure are required.
% Since we are particularly interested in the LOF responses arising from periodic isosceles triangular motifs, a 3D model system relatively realistic in terms of fabrication is chosen for this study, which is illustrated schematically in Figure~\ref{fig:scheme}(a) along with the definition of our coordinate system.
% LOFs are calculated from diffraction efficiencies and simulated by using the rigorous coupled-wave analysis (RCWA) method, a Fourier modal method proficient to calculate optical forces and diffraction efficiencies for layered periodic structures \cite{RCWA}.
% Benefit from the speed for LOF calculation, we employs the Bayesian optimization approach hopin
\subsection{Model System}
A schematic view of the motif model and the choice of coordinate system is depicted in Figure~\ref{fig:scheme}(a).
The motif is planar and 2D periodic in the x- and y-directions with a periodicity given by $\Px=\qty{0.950}{\um}$ and $\Py=\qty{0.600}{\um}$.
These values were chosen to limit the orders of diffraction observed (see below) \cite{MetaV}.
The z-direction is defined along the surface normal of the motif.

The triangular motif is designed to be an isosceles triangle and is oriented in a way such that the whole structure contains a mirror plane perpendicular to the y-axis, which simplifies the system without sacrificing its versatility. 
The in-plane shape of the triangle is uniquely described by two geometric parameters, the widths in the x- and y-directions, denoted by $\wx$ and $\wy$, respectively.
Due to practical concerns, that the infinitely sharp triangle corners in an ideal triangle are not feasible for fabrication and may lead to field singularities, all of the triangle corners are rounded by a circle with radius $\rm R=\qty{5}{nm}$.
A top view of the triangle motif and the definition of key geometric parameters are illustrated in Figure~\ref{fig:scheme}(b).

In line with experimental conditions commonly used for similar devices \cite{MetaV}, the entire motif structure is immersed in water and constructed of three layers: 
a \qty{0.400}{\um} thick layer of SiO\textsubscript{2}, a motif layer consisting of the poly-Si isosceles triangle with a thickness of h=\qty{0.450}{\um} embedded in SiO\textsubscript{2}, and a final SiO\textsubscript{2} layer encapsulating the total device thickness to \qty{1}{\um}. 
Both the substrate and the cap layer are homogeneous, whilst the motif layer is homogeneous in the z-direction and inhomogeneous in the x- and y-directions.
The refractive indices of water, SiO\textsubscript{2}, and poly-Si at the wavelength of our incident light (\qty{1.064}{\um}) are taken as 1.33, 1.45, and 3.45, respectively.

To ensure the geometric asymmetry provided by the triangle motif is the only asymmetry in the optical system, the light source is chosen as a normally incident plane wave with a wavelength in free space of $\lambda=\qty{1.064}{\um}$, i.e. a frequency of $\rm f=\qty{281.76}{THz}$, propagating along the z-direction, linearly polarized in the x-direction.
As a result, the whole system is symmetric in the y-direction and asymmetric in the x-direction.
Thus, the only lateral component remaining for the total optical force $\Ftot$ is given by its x-component $\Fx$; the y-component of the optical force $\Fy$ is exactly zero by symmetry.

\subsection{Optical Force Calculation}

In contrast to direct Maxwell-stress-tensor methods, which require detailed knowledge of the nearfield distributions, the optical forces in this work were obtained indirectly from the farfield diffraction efficiencies, resulting in a substantially lower computational cost.
We used a custom Python toolkit 
% (Python v. 3.9.18, Numpy v. 1.24.3, Scipy v. 1.13.3) 
to perform rigorous coupled-wave analysis (RCWA) simulations\footnote{Code available at https://github.com/hIghwAvE314/RCWA}.
RCWA is a Fourier modal method that directly simulates the flow of light under the approximation of plane wave expansion, which has been proven to be computationally efficient for optical force and diffraction efficiency calculation for layered periodic structures \cite{RCWA}.
Details of the conversion from diffraction efficiencies (direct results of a standard RCWA routine) to optical forces are described below.

Due to the periodic boundary condition, the periodic motif interacts with the light like a diffractive grating.
Given the choice of the incident wavelength $\lambda=\qty{1.064}{\um}$, refractive index of the medium n=1.33, the periods in x- and y-directions $\Px=\qty{0.950}{\um}$ and $\Py=\qty{0.600}{\um}$, only the 0 and $\pm 1$ orders of diffraction are allowed in the x-z plane while diffraction orders higher than 0 (the central order) are prohibited in the y-z plane.  

The fractions of energy distributed in each diffraction order on either the reflection side or the transmission side of the device---also known as the diffraction efficiencies (DEs)---are denoted as $T^-, T^0, T^+, R^-, R^0, R^+$, respectively (see figure~\ref{fig:scheme}c).
According to the conservation of momentum and energy, the optical force can be calculated from knowledge of these DEs.

For computational simplicity, optical forces in this paper are calculated in a dimensionless reduced unit, which is equivalent to the optical forces in the physical units divided by a factor of $P/c$, with $P$ and $c$ being the power of light interacting with the device and the speed of light in free space, respectively.
One of the benefits of working within such a dimensionless unit system is that it automatically excludes the linearity factor of the power of light, and focuses on the interaction between light and the device only.
Suppose the device is illuminated by light with a total power of \qty{1}{mW} per unit cell, assuming all the light is interacting with the device, a LOF in the reduced unit of 1 corresponds to a LOF of \qty{3.34}{pN}.

The optical force exerted on the device is thereby calculated by
\begin{align}
    \mathrm{F_x} &= - \left[(R^+ + T^+) - (R^- + T^-)\right] \sin{\alpha}~,\\
    \mathrm{F_z} &= 1 - (T^0-R^0) - \left[(T^+ + T^-) - (R^+ + R^-)\right]\cos{\alpha}~,
\end{align}
where $\alpha = \sin^{-1}(\dfrac{\lambda}{n\Px}) = 57.36^\circ$ is the diffraction angle of the +1 order.
Here, the x- and z-components of optical force, $\Fx$ and $\Fz$, correspond to a LOF and a forward force (radiation pressure), respectively.
We note that the theoretical upper limit of $\Fx$ in this dimensionless unit system is 0.842, which correspond to all light being diffracted into the $R^-$ and $T^-$ channels (or the two positive channels for negative $\Fx$).
The upper limit of $\Fz$ is 2, corresponding to the case of total reflection with all light being diffracted into the $R^0$ channel.

%----------------------------------------------------------------------------------------
% \section{Result \& Discussion}
\section{Bayesian optimization \label{sec:BO}}

In this section, we aim to probe the maximum achievable LOFs in the triangular motif systems.
While the broken mirror symmetry along the x-direction enables the generation of a nonzero lateral component of optical force ($\Fx$), the magnitude of $\Fx$ may not be inherently sufficient to overcome environmental resistances or fluctuations (e.g., hydrodynamic frictions, thermal fluctuations, and Brownian motions).

\begin{table}
    \centering
    \begin{tblr}{ cells={valign=m, halign=c}, colspec={QQQQQQ},  hlines, vlines, vline{2,4,6}={dashed}}
        \SetCell[c=2]{} Group A & & \SetCell[c=2]{} Group B & & \SetCell[c=2]{} Group C & 
    \\
        \SetCell[c=2]{}
        \begin{tikzpicture}[scale=0.45,baseline={(current bounding box.center)}]
            \draw [fill=gray!20] (0,0cm)rectangle(9.50,6.00cm);
            \draw [rounded corners=0.05cm, fill=gray!80] (0.315,1.685cm) -- (0.315,4.315cm) -- (9.185,3.0cm) -- cycle;
            \node[draw] at (8.0,5.0cm) {A1};
        \end{tikzpicture}
        & &
        \SetCell[c=2]{}
        \begin{tikzpicture}[scale=0.45,baseline={(current bounding box.center)}]
            \draw [fill=gray!20] (0,0cm)rectangle(9.50,6.00cm);
            \draw [rounded corners=0.05cm, fill=gray!80] (1.125,0.405cm) -- (1.125,5.595cm) -- (8.375,3.0cm) -- cycle;
            \node[draw] at (8.0,5.0cm) {B1};
        \end{tikzpicture} 
        & &
        \SetCell[c=2]{}
        \begin{tikzpicture}[scale=0.45,baseline={(current bounding box.center)}]
            \draw [fill=gray!20] (0,0cm)rectangle(9.50,6.00cm);
            \draw [rounded corners=0.05cm, fill=gray!80] (0.385,1.855cm) -- (0.385,4.145cm) -- (9.115,3.0cm) -- cycle;
            \node[draw] at (8.0,5.0cm) {C1};
        \end{tikzpicture}  
        & \\
        {$\wx = \qty{0.887}{\um}$\\$\wy=\qty{0.263}{\um}$} &{$\Fx=\num{ 0.484}$\\$\Fz=\num{1.150}$} %A1
        &
        {$\wx = \qty{0.725}{\um}$\\$\wy=\qty{0.519}{\um}$} &{$\Fx=\num{+0.319}$\\$\Fz=\num{0.485}$} %B1
        &
        {$\wx = \qty{0.873}{\um}$\\$\wy=\qty{0.229}{\um}$} &{$\Fx=\num{+0.332}$\\$\Fz=\num{0.641}$} %C1
    \\
        \SetCell[c=2]{}
        \begin{tikzpicture}[scale=0.45,baseline={(current bounding box.center)}]
            \draw [fill=gray!20] (0,0cm)rectangle(9.50,6.00cm);
            \draw [rounded corners=0.05cm, fill=gray!80] (1.415,1.185cm) -- (1.415,4.815cm) -- (8.085,3.0cm) -- cycle;
            \node[draw] at (8.0,5.0cm) {A2};
        \end{tikzpicture}
        & &
        \SetCell[c=2]{}
        \begin{tikzpicture}[scale=0.45,baseline={(current bounding box.center)}]
            \draw [fill=gray!20] (0,0cm)rectangle(9.50,6.00cm);
            \draw [rounded corners=0.05cm, fill=gray!80] (1.13,0.46cm) -- (1.13,5.54cm) -- (8.37,3.0cm) -- cycle;
            \node[draw] at (8.0,5.0cm) {B2};
        \end{tikzpicture}
        & &
        \SetCell[c=2]{}
        \begin{tikzpicture}[scale=0.45,baseline={(current bounding box.center)}]
            \draw [fill=gray!20] (0,0cm)rectangle(9.50,6.00cm);
            \draw [rounded corners=0.05cm, fill=gray!80] (1.035,1.795cm) -- (1.035,4.205cm) -- (8.465,3.0cm) -- cycle;
            \node[draw] at (8.0,5.0cm) {C2};
        \end{tikzpicture}
        & \\
        {$\wx = \qty{0.667}{\um}$\\$\wy=\qty{0.363}{um}$} &{$\Fx=\num{-0.339}$\\$\Fz=\num{1.156}$} %A2
        &
        {$\wx = \qty{0.724}{\um}$\\$\wy=\qty{0.508}{\um}$} &{$\Fx=\num{-0.259}$\\$\Fz=\num{1.463}$} %B2
        &
        {$\wx = \qty{0.743}{\um}$\\$\wy=\qty{0.241}{\um}$} &{$\Fx=\num{ 0.328}$\\$\Fz=\num{0.597}$} %C2
    \end{tblr}
    \caption{
    Representative structures giving locally maximal LOF obtained from Bayesian optimizations. 
    Group A represents two optimal structures observed with (A1) positive $\Fx$ and (A2) negative $\Fx$.
    % LOF operating in the positive (A1) and negative (A2) direction.
    Group B demonstrates two visually nearly identical structures with distinct and opposite optical force responses.
    Group C shows two visually different structures with similar optical forces.}
    \label{tab:BO}
\end{table}

To determine whether the LOF can potentially be useful, geometry optimizations to maximize the magnitude of $\Fx$ were required. 
While varying the geometric parameters $\wx$ and $\wy$, other parameters such as the wavelength, the thickness, and the radius of rounded corners were kept unchanged. 
Here, the mapping from geometric parameters $\wx$ and $\wy$ to $\Fx$ was treated as a computationally expensive black-box multivariate function with slight stochastic noise (approximately 1\% level), which captures numerical inaccuracies accumulated during simulations.
The geometry optimization under such a treatment fits well with the scope of Bayesian optimization, where the data retrieved from an objective function evaluation may bear uncertainty making it preferable to perform fewer function evaluations due to its high cost.

The Bayesian optimizations were implemented with an open-sourced package \cite{B/O}, and employed domain reduction \cite{B/O-soft} to accelerate convergence.
LOF simulations were wrapped up as a function taking $\wx$ and $\wy$ as input and returning $\Fx$ or $-\Fx$ depending on the desired sign.
This is to ensure the directionality of LOFs obtained from optimizations.
We used the ``upper confidence bound" (UCB) acquisition function \cite{B/O} to suggest the next point in the parameter space to probe.
The acquisition function is defined as
\begin{equation}
    \vb{x}^* = \mu(\vb{x}) + \kappa \sigma(\vb{x})~,
\end{equation}
where $\mu(\vb{x})$ and $\sigma(\vb{x})$ are the mean and standard deviation predicted by Gaussian process regression (GPR) with a Matern kernel and a 1\% noise-level (also known as the $\alpha$ value in GPR), fitted to data from earlier steps.
$\kappa$ is a tunable parameter determining the confidence level for the prediction, and it is initialized at 2.576 (corresponding to a 99\% confidence level) and multiplied by 0.99 at each iteration, resulting in a gradual decay.
The optimization started with 20 independent random searches within the parameter space to build up unbiased priors, and then performed Bayesian optimization iterations until no larger LOF had been observed for 30 steps, which was considered as the convergence criterion.

Out of 20 repeated numerical experiments with different random seeds, 6 representative structures giving locally maximal LOF are summarized in Table~\ref{tab:BO}, forming 3 groups (A, B and C) demonstrating characteristic behaviors that were typically observed.
The two structures in Group A reveal the first counterintuitive phenomenon: Strong LOFs can be generated in both positive (A1: +0.484) and negative (A2: -0.339) directions despite the same orientation of the triangles. 
This contradicts the conventional geometric intuition that the fixed orientation of structural asymmetry in triangle A1 and A2 would typically suggest consistent force directionality.

Other surprising phenomena emerge when comparing Group B and C.
Structures B1 and B2 appear visually identical, differing only by $\sim\qty{10}{nm}$ (2\%) in $\wx$, yet generate opposing LOFs (+0.319 vs -0.259) and show substantial variation in the forward force component $\Fz$ (0.485 vs 1.463).
Conversely, structures C1 and C2 exhibit striking visual differences ($\Delta\wx\approx\qty{0.130}{\um}$) but produce nearly identical optical force responses in both lateral ($\Fx \sim 1\%$ different) and forward components ($\Fz \sim 5\%$ different).

These observations suggest two distinct geometric regimes for LOF generation: (1) a highly unstable regime where minimal geometric variations (as in Group B) induce dramatic LOF magnitude changes or direction reversals; (2) a stable regime where forces remain stable across substantial geometric differences (as in Group C).
The two distinct regimes in structure-force relationships have important implications for optical manipulation systems, suggesting that both precision control of optical forces or tolerance to geometric variation can be achieved through appropriate structural design.

\section{Landscapes in parameter space \label{sec:landscape}}

The result of the Bayesian optimization approach suggests that the optical force variation across the parameters space are complex, offering a rich and mixed behavior, which was not appreciated by only focusing on the maximal values.
Therefore, to visually demonstrate the behavior observed during Bayesian optimizations and their associated LOF behaviors, we systematically mapped the parameter space, obtaining a landscape of LOFs across various combinations of geometric parameters $\wx$ and $\wy$.
\begin{figure}
    \centering
    \includegraphics[width=\linewidth]{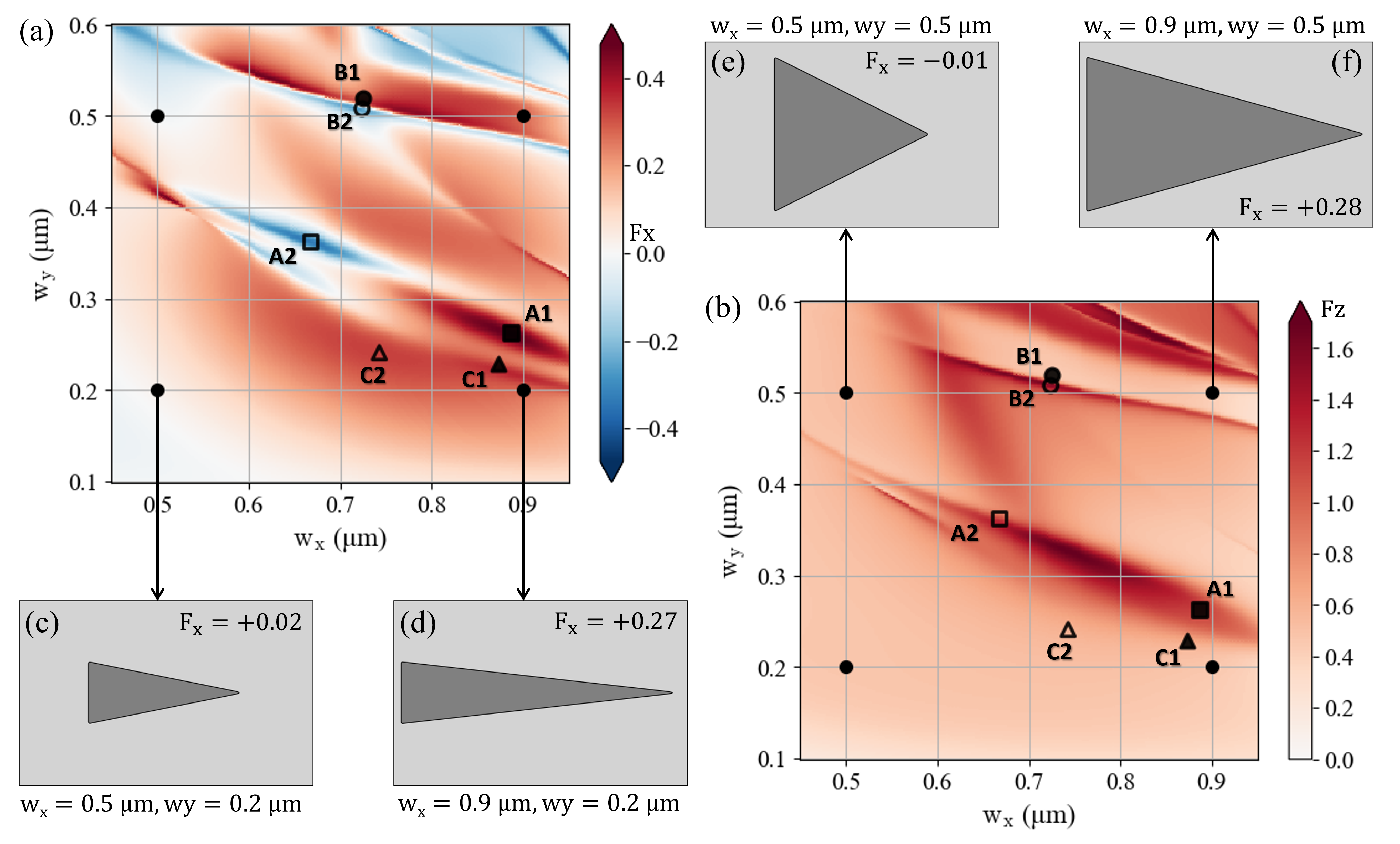}
    \caption{
    (a) A contour plot of LOF. The color at each point represents the value of $\Fx$ of the structure with corresponding $\wx$ and $\wy$. Red and blue indicate the direction of $\Fx$ being positive or negative, respectively. 
    (b) A contour plot of the forward component of optical force, $\Fz$, for completeness.
    (c-f) Four structures along with their corresponding $\wx$, $\wy$ and $\Fx$ values are labeled aside as examples. 
    The motif thickness in this figure is kept fixed at h=\qty{0.450}{\um}.
    The six structures summarized in Table~\ref {tab:BO} are also marked on the figure using the same labels.
    \label{fig:mapping}
    }
\end{figure}

Figure~\ref{fig:mapping}(a) demonstrates the landscape of $\Fx$, with $\wx$ scanning from \qty{0.450}{\um} to \qty{0.950}{\um} and $\wy$ from \qty{0.100}{\um} to \qty{0.600}{\um} in steps of \qty{5}{nm}.
The color at each point indicates the value of $\Fx$ for the corresponding structure, with red for the positive direction and blue for the negative direction.
The faint white-ish color suggests a nearly zero LOF.
Four example structures with their parameters and visual appearances are displayed to the side (panel c-f) for better illustration.
The six structures reported in Table~\ref{tab:BO} are also labeled in the figure.
The landscape of $\Fz$ is displayed in panel (b) for completeness.

The upper limits of $\wx$ and $\wy$ are determined by the periodicity of the structure $\Px$ and $\Py$.
When the values of $\wx$ and $\wy$ become smaller than the displayed ranges, \qty{0.45}{\um} and \qty{0.1}{\um}, respectively, the asymmetry in the triangle motif quickly diminishes, and the resulting LOF values become negligible as seen in the bottom left corner.  

% The $\Fx$ landscape shown in Figure~\ref{fig:mapping}(a) displays a pattern of stable zones and switching regions.
From the $\Fx$ landscape shown in Figure~\ref{fig:mapping}(a), it is noticeable that $\Fx$ changes gradually in some regions while changing dramatically in other regions.
These are thereby referred to as the stable zones and switching bands, respectively.
More specifically, stable zones are regions where $\Fx$ has predominantly positive (red) or negative (blue) values, varying gradually toward localized extrema.
Between these zones the switching bands typically appear, in which the value of $\Fx$ changes rapidly between positive and negative values, appearing as sharp but continuous color transitions (e.g., the white-ish regions between red and blue regions).
The interplay between stable zones and switching bands offers a rich behavior for further exploration.

\begin{figure}
    \centering
    \includegraphics[width=\linewidth]{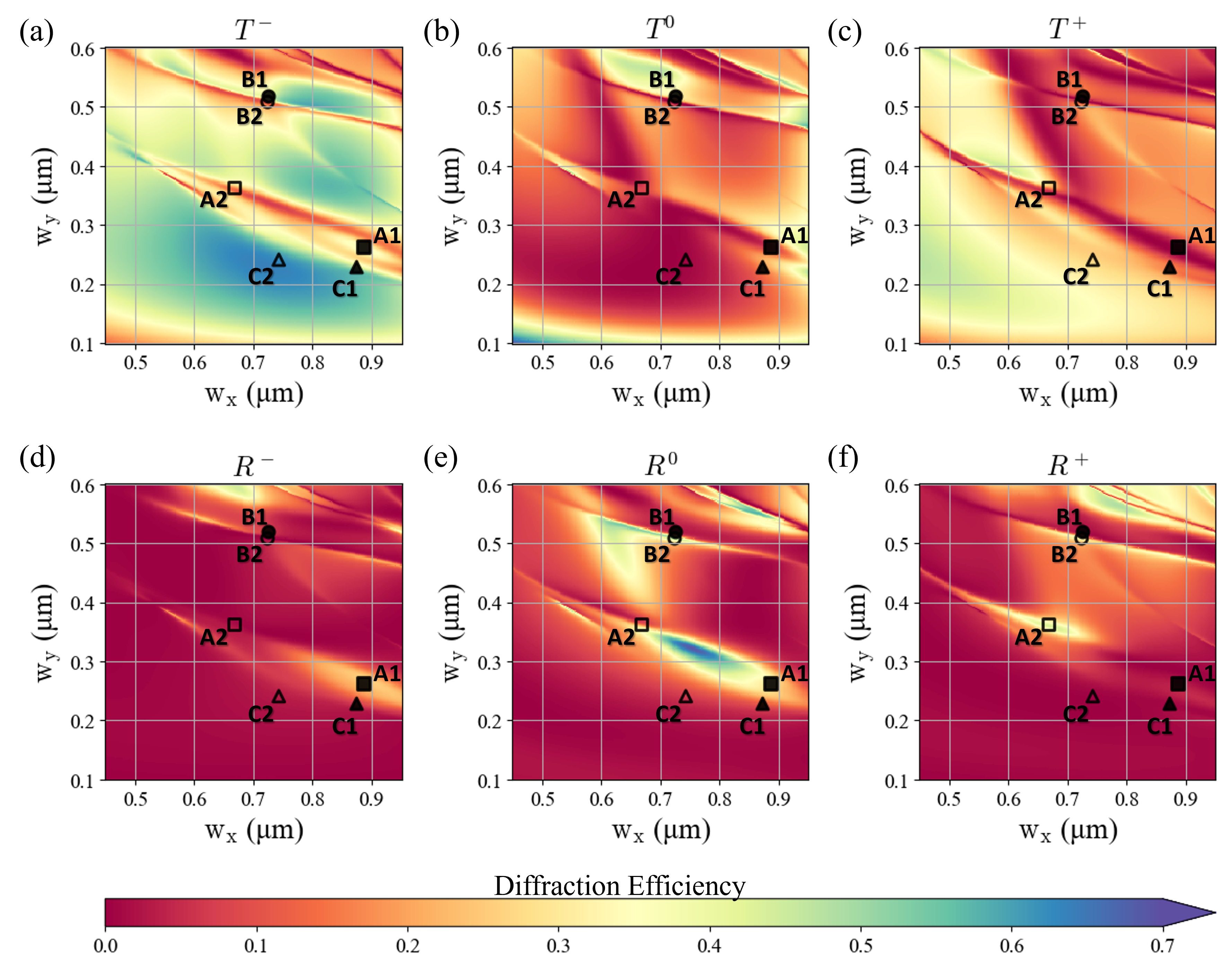}
    \caption{
    Diffraction efficiency contour plot of (a) $T^-$, (b) $T^0$, (c) $T^+$, (d) $R^-$, (e) $R_0$, and (f) $R^+$ diffraction channels, as defined in Figure~\ref{fig:scheme}(c). 
    % Landscapes of (g) $\Fx$ and (h) $\Fz$ for completeness. 
    The six structures summarized in Table~\ref{tab:BO} are also marked on the figures for reference.
    \label{fig:DEs}
    }
\end{figure}

To resolve whether the $\Fx$ generation mechanisms are consistent within each stable zone and vary across zones, we analyze the DEs for each of the transmitted and reflected diffraction orders, as demonstrated in Figure~\ref{fig:DEs}.
At each corresponding point in these figures, the values of the DEs sum to 1 due to energy conservation, since there is no absorption in the system.  
If the values of DEs in certain diffraction channels are more dominant, other diffraction channels will diminish.
The six structures summarized in Table~\ref{tab:BO} are marked in the figure for reference and comparison.
% We also plot the variation in $\Fx$ and $\Fz$ along with the six structures summarized in Table~\ref{tab:BO} for reference, comparison, and completeness.

These landscapes combined provide physical insights into LOF generation and reveal systematic shape-dependent behaviors. 
The DE color presented for each channel, visualizes whether certain diffraction orders are more enhanced or avoided, which helps to explain the mechanism of the optical force generation from a farfield perspective.
A stable zone in the $\Fx$ landscape typically corresponds to a relatively uniformed DE color combination in either $R^+$, $R^-$, $T^+$, or $T^-$ channels, which demonstrates a persistent diffraction-order dominance.
In contrast, in the switching bands of the $\Fx$ landscape where the value of $\Fx$ changes rapidly, such a dominance in diffraction order may also alter rapidly, typically accompanied with a higher value of DE in the $R^0$ or $T^0$ channel.
% highlighting physical significance in distinguishing switching bands from stable zones in the $\Fx$ landscape.
Furthermore, when more light is diffracted to the reflection side, the structure will experience a higher magnitude of $\Fz$, and lower $\Fz$ if more light is diffracted to the transmission side.

Therefore, these contour plots demonstrate a fundamental representation of LOFs that enhances the understanding of the LOF generation due to a given structure as well as LOF classifications.
Take the six structures from the Bayesian optimization results as examples.
The two structures of Group A (with square markers) belong to two distinct stable zones in the $\Fx$ landscape as shown in Figure~\ref{fig:mapping}(a), which demonstrate completely different LOF generation mechanisms (see below).
Structure A1 with a positive $\Fx$ shows moderate DEs in the $R^0$ (0.28), $R^-$ (0.23) and $T^-$ (0.39) channels and low DEs ($\lesssim 0.05$) in all other three channels, indicating a dominant -1 order diffraction on the transmission side and minimal +1 order diffraction on the reflection side. Overall diffracted light for structure A1 is going to the negative direction.
Structure A2, on the other hand, demonstrates a rather high DE in the $R^+$ (0.43) channel, relatively low DE in the $R^0$, $T^+$, and $T^-$ channels ($\sim 0.15$), and minimal DE in the $T^0$ (0.07) and $R^-$ (0.01) channels.
This exhibits a +1 order diffraction dominance on the reflection side and roughly evenly distributed transmissive diffraction (slightly less on the 0 order), rendering more light being diffracted into the positive direction, which explains the negative $\Fx$. 
Both structures exhibits a stronger reflection than transmission, supporting their similar large $\Fz$ values ($\Fz \sim 1.15$).

Structures B1 and B2 (with circle markers) are separated by a sharp switching band.
The B1 structure with a positive $\Fx$ demonstrates a minimal level of DE values in all three reflection channels (all $< 0.07$), while the transmission side DEs are moderately low on the $T^+$ channels (0.11) and high on the $T^0$ (0.32) and $T^-$ (0.43) channels, suggesting a positive $\Fx$ and a small $\Fz$.
Structure B2, however, appears completely different.
It shows strong DEs on the $R^0$ (0.44) and $R^+$ (0.31) channels, moderately low on the $T^-$ and $T^+$ channels ($\sim 0.11$) and minimal on the $R^-$ and $T^0$ channels ($\sim 0.01$).
This reflects the relatively small negative $\Fx$ and large $\Fz$ observation. 
The magnitude of $\Fx$ is relatively small ($< 0.3$) because the diffracted light is spread across more channels, with a significant amount of light going to both positive and negative directions.
% is more diverged in terms of directionality with a standing-out positive direction channel ($R^+$).
Comparing B1 and B2, it is clear that the distinct LOF that change in sign must be related to the two structures being either side of the sharp switching band, which will be further investigated in later sections.

Structures C1 and C2 (with triangle markers) belong to the same stable zone despite the large shape variation.
This stable zone exhibits a very high DE level on the $T^-$ channel (0.5-0.6) and moderately low level on the $T^+$ channel (0.1-0.2), with other channels being minimal or low ($\lesssim 0.1$).
As a result, the structures within this stable zone typically reflect less light than they transmit, and diffract significantly more light to the negative direction than to the positive direction, rendering a positive $\Fx$ and a small $\Fz$ at similar levels.

\section{Observation of Fano-like behaviors \label{sec:spec}}

\begin{figure}[h]
    \centering
    \includegraphics[width=\linewidth]{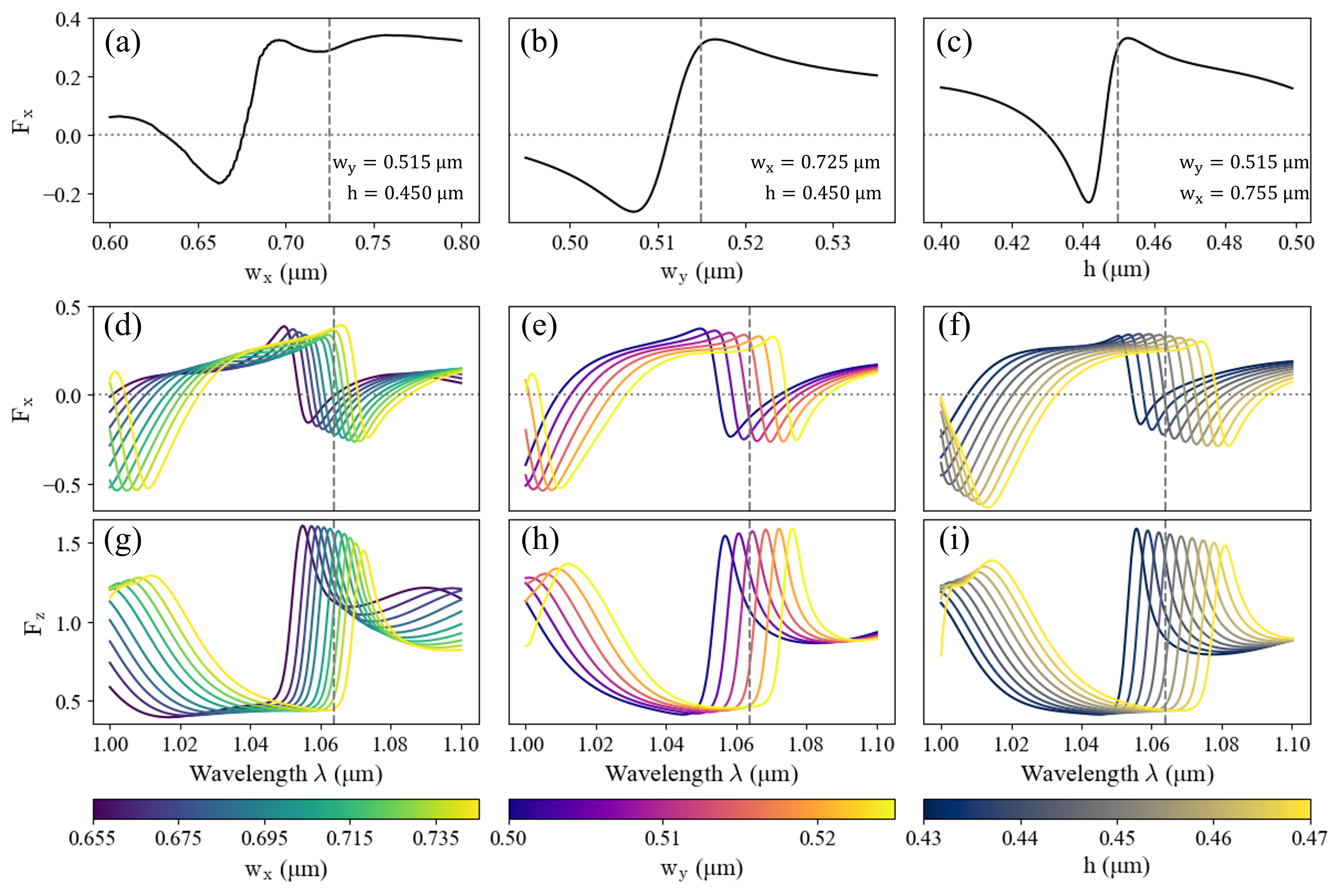}
    \caption{
    Focusing on the structure with $\wx=\qty{0.725}{\um}$, $\wy=\qty{0.515}{\um}$, and triangle height ${\rm h}=\qty{0.45}{\um}$, the $\Fx$ responses are shown with respect to changes in (a) $\wx$, (b) $\wy$, and (c) $\rm h$. The wavelength dependence of (d-f) $\Fx$ and (g-i) $\Fz$ of the three series of structures with a gradually shifted color scheme are displayed, respectively.
    The vertical dashed lines represent the reference structure in (a-c) and the reference incident wavelength in (d-i). The horizontal dotted lines in (a-f) mark $\Fx=0$.
    }
    \label{fig:spec}
\end{figure}

We now examine the optical force behavior within switching bands through systematic parameter exploration. 
A representative structure located between points B1 and B2 ($\wx=\qty{0.725}{\um}, \wy=\qty{0.515}{\um}, {\rm h}=\qty{0.450}{\um}$) was selected as a reference point, since B1 and B2 exhibit characteristic switching band behavior.
Centered at this structure, we individually vary each geometric parameter ($\wx$ and $\wy$), as well as an additional parameter, the triangle thickness h, to analyze their effects on $\Fx$ (Figure~\ref{fig:spec}a-c).
The thickness h is introduced as an external parameter that was not included in the earlier landscape and Bayesian optimization studies.
Including h allows us to test whether the trends identified through variations in $\wx$ and $\wy$ remain consistent when the out-of-plane geometry is changed, thereby helping us evaluate the robustness and potential generality of those trends.
For completeness, we also examine the wavelength-dependent responses of $\Fx$ and $\Fz$ as a function of incident wavelength, centered at $\lambda=\qty{1.064}{\um}$ (Figure~\ref{fig:spec}d-i), along with the effect of varying $\wx$, $\wy$, and h on the force--wavelength dependence.
The vertical dashed lines represent the reference structure in (a-c) and the reference incident wavelength in (d-i).

All three $\Fx$ curves in Figure~\ref{fig:spec}(a-c) show a similar pattern.
$\Fx$ changes sharply near the reference structure but varies little elsewhere.
Outside this transition region, increasing $\wx$, $\wy$, or h causes $\Fx$ to either decrease gradually or remain nearly constant, while within the transition region, it rises rapidly from minimum to maximum.
The wavelength-dependent responses of $\Fx$ (d-f) and $\Fz$ (g-i) demonstrate a monotonic shift in wavelength as the geometric parameters increase.
Meanwhile, the spectral response of each structure contains a similar rapid change region around the reference wavelength, in line with the $\Fx$ curve appearance in (a-c).

Further analysis of the spectral $\Fx$ (d-f) and $\Fz$ (g-i) curves reveals important optical force properties with practical implications. 
The behavior in these ranges provides crucial data relating to fabrication tolerances, while additional spectral features suggest further application possibilities worth exploring, as claimed below.

% Figure~\ref{fig:spec} (d-f) and (g-i) demonstrate $\Fx$ and $\Fz$ spectra for a series of representative structures during changing individual parameters, respectively.
Consider, for instance, the development of a wavelength-tunable bidirectional optical conveyor or lift operating perpendicular to the light propagation direction, where the monotonic force--wavelength response around \qty{1.064}{\um} enables precise directional switching through minor wavelength adjustments.
These spectral curves in Figure~\ref{fig:spec}(d-i) directly support this application by identifying the spectral characteristics of $\Fx$ and the corresponding wavelength region with monotonic $\Fx$ responses for each corresponding structure. 
They also suggest an optimal fabrication window for each parameter in which the target wavelength is included in their monotonic region:
a \qty{40}{nm} window for $\wx$ (from \qty{0.695}{\um} to \qty{0.735}{\um}), a \qty{10}{nm} window in $\wy$ (from \qty{0.505}{\um} to \qty{0.515}{\um}), and a \qty{20}{nm} window for h (from \qty{0.450}{\um} to \qty{0.470}{\um}).
These quantitative guidelines simultaneously establish fabrication tolerances.
Beyond this specific application, these spectra contain additional features that may enable other novel functionalities, warranting further investigation.

While our primary focus remains on switching bands, structures in stable zones (such as Group A and C) present complementary advantages for applications where stability outweighs sensitivity needs.
These structures maintain consistent optical force responses across geometric variations, exhibiting strong tolerance to both fabrication errors and operational deformations.
This stability profile suggests their potential for devices requiring reliable performance under complex or variable conditions, where maintaining steady optical forces is essential. 

The observed force spectra also reveal an important connection to Fano resonances, a fundamental photonic phenomenon where high-Q eigenmodes interfere with propagating light to produce characteristic asymmetric lineshapes \cite{Fano, highQvdW}.
These distinctive dip-and-peak profiles appear in both $\Fx$ and $\Fz$ spectra.
While optical forces are not intrinsic resonance quantities, they directly link to conventional resonance responses. 
Specifically, $\Fx$ quantifies the extent of asymmetry in the resonant field distributions (via diffraction profiles), whereas $\Fz$ measures transmittance/reflectance.
This connection explains why both force components display Fano-like spectral features in Figure~\ref{fig:spec}, as the Fano-resonance frequency shifts systematically with geometric parameters.
Similar dip-and-peak features emerge when tracking $\Fx$ at \qty{1.064}{\um} across geometric variations (Figure~\ref{fig:spec}a-c), even though these parameter-space patterns differ from true resonances as they map structural parameters rather than wavelength dependencies.
This dual manifestation highlights the strong connection between the switching band patterns and Fano resonance physics.

Notably, while our earlier studies in Section~\ref{sec:BO} and \ref{sec:landscape} focused only on $\wx$ and $\wy$, the inclusion of h reveals identical switching patterns and Fano-like spectral features. 
This consistency across all three geometric dimensions suggests that a universal principle governs the observed optical force transitions, one that likely transcends our specific triangular system. 
The phenomenon's robustness implies its physical origin is fundamental, likely rooted in the interference physics shared by Fano resonances and optical force generation.
Understanding this principle could reshape design strategies for functional optically-driven materials and devices.
By exploiting the coupling between geometric parameters and resonant light-matter interactions, future devices might achieve tailored force responses across diverse platforms, from metasurfaces to photonic integrated circuits.

While the Fano-like lineshapes observed in the optical force spectra hint at an underlying resonance mechanism, their origin requires further analysis.
We therefore proceed with an eigenfrequency study in the following section.
% In particular, Fano resonance is typically considered as an discrete quantum state interfering with a continuum band in photonic crystals.

\section{Eigenfrequency analysis \label{sec:eigen}}

The previous section revealed optical forces exhibiting dip-and-peak asymmetric lineshapes resembling Fano resonances.
Although this hints at resonant interference effects, confirming whether these features genuinely arise from Fano resonances---and whether this relationship holds across multiple switching bands---necessitates a deeper investigation of the underlying eigenfrequencies of the system.

The origin of Fano resonances is typically considered as a discrete localized quantum state interfering with a continuum band in photonics. 
One of the famous examples is Mie scattering, or Mie resonance, in which a spherical Mie particle (sometimes a cylinder) interacts with the incident light and demonstrates a series of Fano resonance characteristics in its scattering spectrum.
These resonances occur when the frequency of the incident light matches the frequency of an eigenmode of the Mie particle, such as the electric/magnetic dipolar mode or quadrupole mode \cite{Fano}.

\begin{figure}
    \centering
    \includegraphics[width=0.8\linewidth]{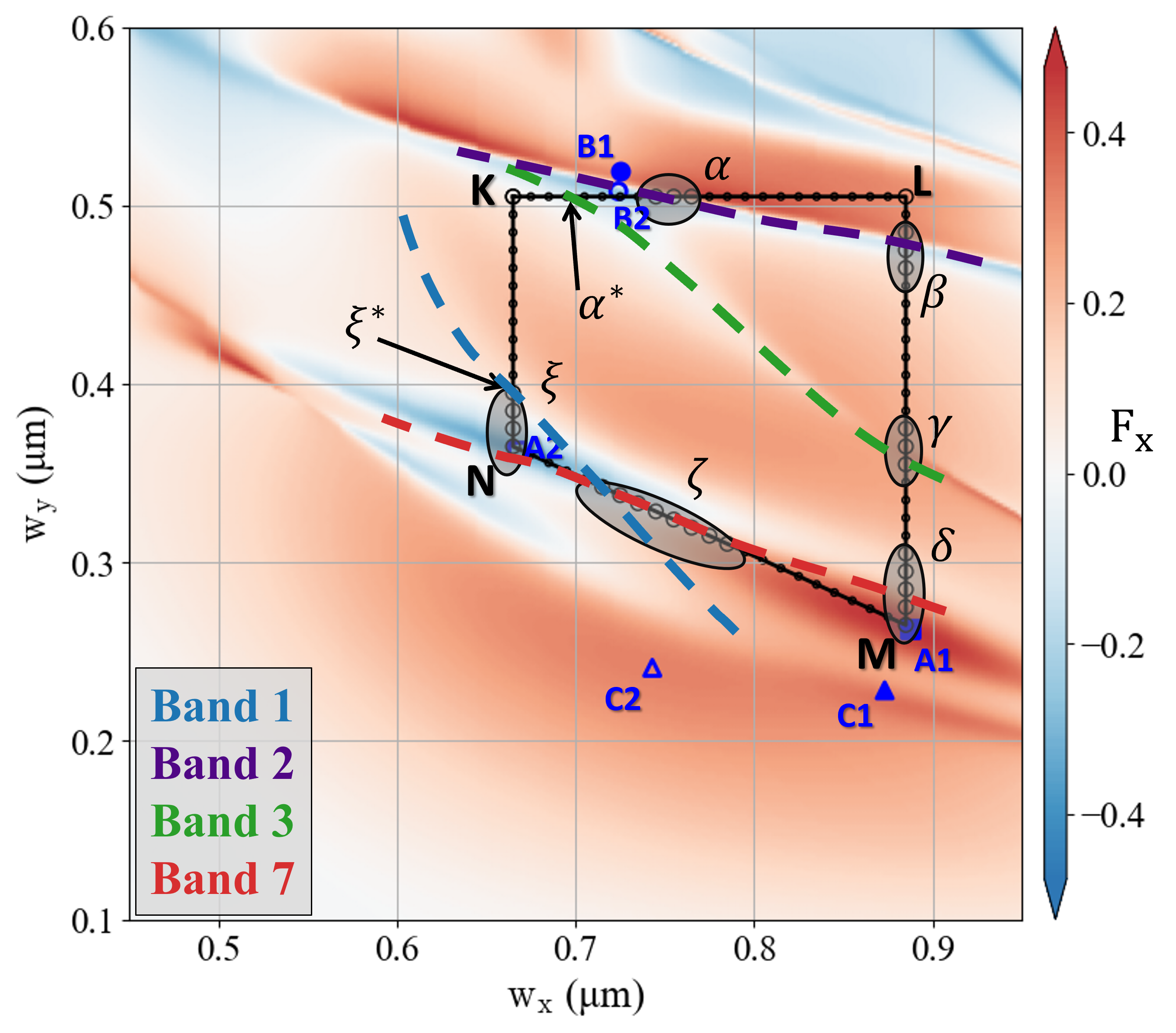}
    \caption{
    A selected loop K-L-M-N-K in the $\Fx$ landscape, with segments crossing switching bands circled in black and denoted by $\alpha,\,\beta,\,\gamma,\,\delta,\,\zeta,\,~\text{and}~\xi$, respectively.
    The six structures summarized in Table~\ref{tab:BO} are denoted by blue markers and labels for reference.
    Four switching bands are sketched and labeled with dashed lines for reference.
}
    \label{fig:eigen_a}
\end{figure}

To verify if our case can be explained similarly, we analyze structures along the parameter-space loop K-L-M-N-K in Figure~\ref{fig:eigen_a}, which is designed to probe multiple stable zones and switching bands.
The loop's trajectory in parameter space enables a systematic examination of how eigenmode evolution correlates with the observed force spectra.

The four corner points of the loop (K, L, M, N) are selected such that points M and N lie within stable yet narrow zones exhibiting the strongest $\Fx$ response, as seen in Group A (Table~\ref{tab:BO}).
Points K and N, as well as L and M, differ only in their $\wy$ values, while points K and L share the same $\wy$.
The loop intersects switching bands at six segments labeled $\alpha$, $\beta$, $\gamma$, $\delta$, $\zeta$, and $\xi$, circled in black.
For example, the extended switching band that separates structures B1 and B2 (band 2 on Figure~\ref{fig:eigen_a}), intersects the line segment K-L.
These circled segments are of particular interest as each may represent a Fano resonance candidate.

\begin{figure}
    \centering
    \includegraphics[width=\linewidth]{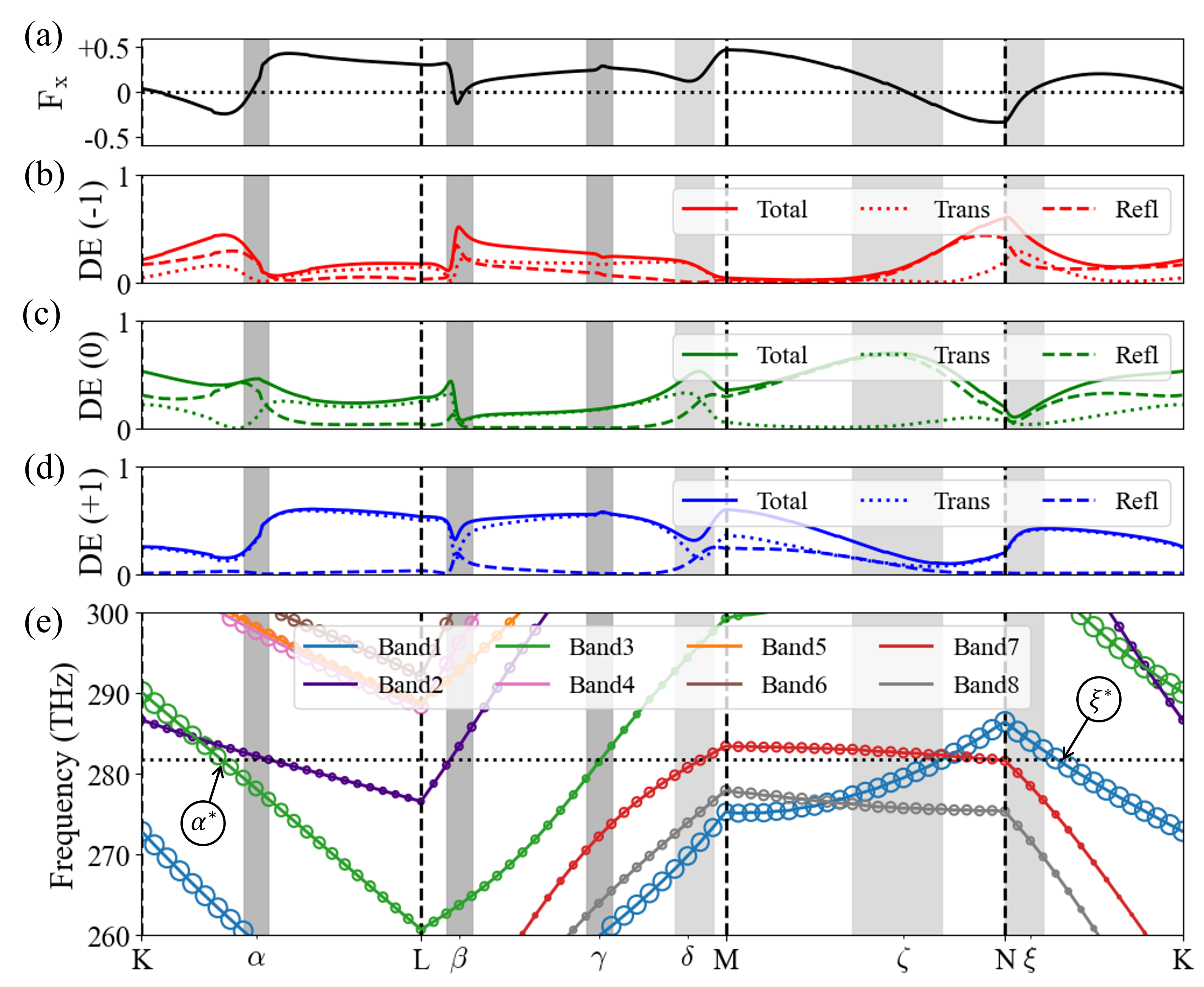}
    \caption{
    (a) Variation of $\Fx$ around the loop shown in Figure~\ref{fig:eigen_a}.
    Variation of DEs around the loop for (b) -1, (c) 0, and (d) +1 diffraction order.
    Dotted and dashed lines denote $T^-/T^0/T^+$ and $R^-/R^0/R^+$, respectively, while solid lines represent total DEs with both sides considered.
    (e) Eigenfrequencies close to the incident light frequency for structures around the loop. Eigenmodes with polarization mismatched with the incident light are neglected.
}
    \label{fig:eigen_b}
\end{figure}

Figure~\ref{fig:eigen_b}(a-d) shows $\Fx$ and DEs for each diffraction order along the loop, with K, L, M, N marked by black dashed lines, while circled segments of interest are shaded with gray background.

When comparing the gray, shaded segments of these $\Fx$ and DE curves (Figure~\ref{fig:eigen_b}a-d), segments $\alpha$, $\beta$, and $\gamma$ exhibit Fano-like asymmetric lineshapes in geometric parameter space, whereas segments $\delta$, $\zeta$, and $\xi$ do not show a complete dip-and-peak profile. 
We note that the Fano-like dip-and-peak profile for segment $\gamma$ appears relatively small.
However, these profile deviations do not necessarily contradict the Fano resonance interpretation and can be explained by further examination of the corresponding $\Fx$ behavior in Figure~\ref{fig:eigen_a}. 
Much of the explanation relates to the fact that the section chosen through parameter space, K-L-M-N-K, intersects the dip-and-peak profiles of the Fano resonances obliquely, or only partially, thus distorting the shape of the profile in Figure~\ref{fig:eigen_b}(a-d).
For example, since the points M and N are chosen to be local extrema and each represents a distinct stable zone of $\Fx$, the expected full dip-and-peak Fano-like lineshapes are consequently truncated---either a region before the dip (off-dip) or after the peak (off-peak), or a region between the dip and the peak (dip-to-peak).
Furthermore, Segment $\zeta$ corresponds to the switching band between the two stable zones represented by points M and N, where $\Fx$ transitions monotonically and relatively gradually.
Extending the segment region of $\zeta$ along the line M-N, as demonstrated in the $\Fx$ curve between M and N (Figure~\ref{fig:eigen_b}a), a peak-to-dip profile appears, supporting the aforementioned truncation effect.
Similarly, segment $\delta$ and $\xi$ only display the off-peak and the off-dip region of the full Fano-like profile.

%Say something about the Q-factor of the corresponding resonance for these three segments are low, so the Fano profile is broader (larger linewidth)

% Meanwhile, segment $\xi$ and $\delta$ display only half of the characteristic dip-and-peak profile (i.e., only the dip or peak).
% as they may arise from low-Q resonances truncated by the choice of points M and N, which themselves represent local extrema (either the dip or peak).
% Extending the analysis beyond these points, as suggested by Figure~\ref{fig:eigen}(a), reveals that $\Fx$ indeed exhibits the full dip-and-peak behavior associated with Fano resonances.

Figure~\ref{fig:eigen_b}(e) demonstrates eigenfrequencies near the incident light frequency (\qty{281.76}{THz}) calculated for structures along the parameter-space loop K-L-M-N-K.
The eigenfrequency simulations were performed using COMSOL Multiphysics with perfect-matching-layer (PML) boundary conditions on the propagation sides to avoid reflection from simulation cell boundaries.
We excluded eigenmodes exhibiting nonphysical field distributions (where fields were concentrated primarily in the perfect-matching-layers rather than within the structures) or symmetry mismatches (specifically, modes that could only be excited by y-polarized light).

The eigenfrequencies were tracked and connected by verifying the consistency of their corresponding eigenmode field distribution.
This approach resembles the construction of band diagrams, with the key difference being that the x-axis represents different structural configuration along the parameter-space loop rather than conventional wavevector values in k-space.

These eigenfrequency bands are numerically labeled from 1 to 8, where the marker size for each mode indicates its Q-factor through an inverse relationship, i.e., larger markers correspond to lower Q and broader bandwidths.
It should be noted that these numerical labels do not reflect the relative spectral positions of the bands.

Clear correspondence is observed between the switching bands and points where eigenfrequencies match the incident light frequency.
Specifically, at each segment of interest, an eigenfrequency band crosses the incident frequency (indicated by a horizontal dotted line):
band 2 at segment $\alpha$ and $\beta$, band 3 at $\gamma$, band 7 at $\delta$, band 1 at $\zeta$, and band 7 at $\xi$.
Aside from these six segments, Figure~\ref{fig:eigen_b}(e) also reveals two hidden segments that intersect with switching bands, $\alpha^*$ and $\xi^*$, as annotated on panel (a) and (f).
These two are too close to other segments to be distinguish solely from the $\Fx$ landscape, while their broad eigenfrequency bandwidth further hinder their visibility.

\begin{figure}
    \centering
    \includegraphics[width=\linewidth]{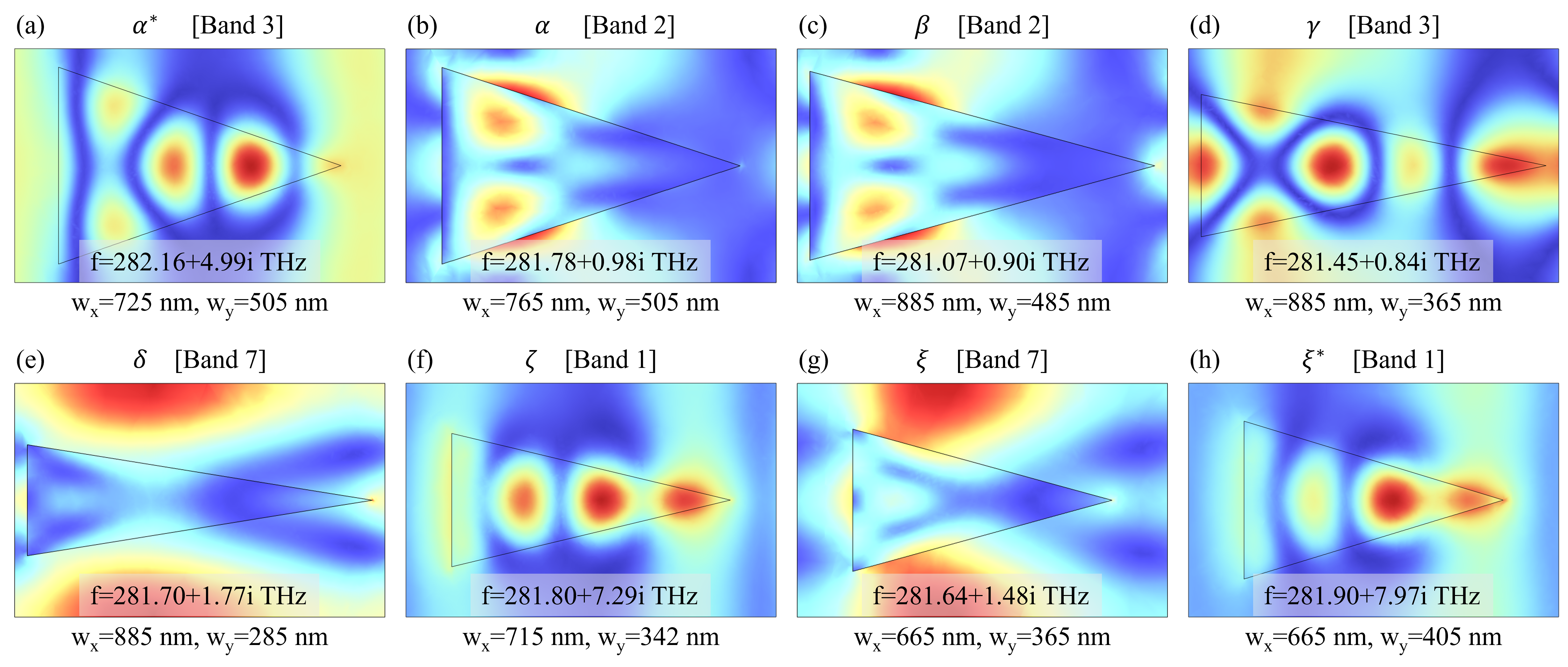}
    \caption{
    \label{fig:field}
    Normalized electric field magnitude at the central XY-plane of the triangle for eight selected structures representing the matching of eigenfrequency with the incident frequency.
    The corresponding band number and the loop position as described in Figure~\ref{fig:eigen_a} and Figure~\ref{fig:eigen_b} are labeled at the top of each panel.
    Associated geometric parameters $\wx$ and $\wy$, as well as the complex eigenfrequency are given at the bottom of each panel.
    }
\end{figure}

As a result, eight structures were selected, representing the intersection of eigenfrequency bands in Figure~\ref{fig:eigen_b}(e) and the incident frequency.
Their typical eigenmode field profiles are shown in Figure~\ref{fig:field}, in the order of occurrence around the loop from $\alpha^*$ to $\xi^*$.
Each panel presents the electric field magnitude at the central XY-plane of the triangle, normalized for better color representations, accompanied with their corresponding geometric parameters $\wx$ and $\wy$, complex eigenfrequency, and band label.
The full 3D profile is not shown.
Nevertheless, it is clear that the fields show high similarity when their band numbers match.

Several important relationships emerge when examining the switching bands, parameter-space segments, and eigenfrequency bands collectively.
For clarity and reference, four switching bands associated with the selected loop are shown schematically on Figure~\ref{fig:eigen_a}.
The sketched lines indicate only their general locations, with colors and labels matching the excitations of the corresponding eigenmodes (panel f).
Segments $\alpha$ and $\beta$ correspond to the same switching band (band 2), as do $\delta$ and $\xi$ (band 7), which can be deduced by the connected whitish bands and adjacent stable zones in these regions of the $\Fx$ plot (Figure~\ref{fig:eigen_a}).
In the case of band 2 ($\alpha$ and $\beta$) the small markers indicate excitation of a high-Q eigenmode ($Q \sim 150$).
Considering that a higher Q value typically indicates a sharper resonance (harder to be excited), this is consistent with the rapid $\Fx$ transition observed in band 2.

As noted above, an additional switching band exists between point K and segment $\alpha$, where $\Fx$ transitions from negative to positive ($\alpha^*$).
The $\Fx$ transition here appears more gradual, partly due to the orientation of the K-L line, and indicates a relatively broad resonance.
From Figure~\ref{fig:eigen_a} and Figure~\ref{fig:eigen_b}(e), this hidden band can be identified as part of the same switching band as $\gamma$, which arises because of the eigenfrequency band 3.
The Q-factor of band 3 demonstrates large variation, decreasing from approximately 200 at $\gamma$ to about 15 at point K, consistent with the broadening of the force profile between those two points.

For segments $\delta$, $\zeta$, $\xi$, and $\xi^*$, the behavior is more complex.
All three switching bands are exceptionally broad, as discussed in the last section (off-peak/dip or dip-to-peak), partly due to the choice of the points M and N at extrema, and partly due to complex eigenfrequency bands coupling and broadening.
Segment $\delta$ corresponds to the excitation of an intermediate-Q mode (band 7, $Q \sim 80$) while its slope is relatively flat compared to the bands 2 and 3, which explains the broadening of $\delta$.
A flatter slope in eigenfrequency band at the incident frequency indicates that this mode continues to be excited at a larger distances in parameter space, so the Fano-like profile will appear stretched or broadened.
Segment $\zeta$ is broadened due to excitation of the same mode (band 7) and an even flatter slope, as well as due to coupling with another low-Q band (band 1, $Q \sim 20$).
% Segment $\xi$ can be explained as the excitation of the low-Q band 1 coupling with band 7.
Segment $\xi$ and the hidden segment $\xi^*$ correspond to the excitation of the intermediate-Q band 7 and the low-Q band 1, respectively.
However, both bands are relatively broad and close to each other, rendering a shifted and deformed $\Fx$ transition appearance in the proximity.

The eigenfrequency band diagram (Figure~\ref{fig:eigen_b}e) can also be used to discuss band formations and intersections.
The hidden segment $\alpha^*$ and the segment $\zeta$ are the only two places along the given loop in Figure~\ref{fig:eigen_a} that demonstrate a merging feature in parameter space, whereas in the band diagram these two correspond to the only two occasions where two potentially excited eigenfrequency bands intersect and interchange.
At the proximity of $\alpha^*$ (upper left), the two distinct switching bands band 2 and band 3 are getting closer and merged into one, which may separate again as they go further.  
This behavior is in line with the band 2 and band 3 intersection near the incident frequency at the proximity of $\alpha^*$.
On the other hand, the near-parallel alignment between line M-N and the $\delta-\xi$ switching band (band 7) matches the minimal variation in band 7's frequency between points M and N.
The intersection of switching band 7 and band 1 happens within the segment $\zeta$, which coincides with the crossing of eigenfrequency band 7 and band 1 in Figure~\ref{fig:eigen_b}(e).
% At the proximity of $\xi$ and point N, band 7 and band 1 ($\xi$-right) intersect the incident frequency imminently but slightly separated.
% Although they are likely coupled and both being excited within this region, the excitation dominance of the two mode may be exchanged, supporting the existence of the two switching bands above and below point N.
% Notably, the switching band below point N is not included in the loop, but it can be tracked as the same switching band intersect segment $\zeta$, confirming that they are likely both band 7 dominated.

The clear correspondence between eigenmodes and the $\Fx$ switching bands (Figure~\ref{fig:eigen_a}) confirms that the Fano-like lineshapes in the optical force spectra are indeed due to the coupling between discrete eigenmodes and propagation.
Moreover, our analysis indicates that the sharpness of the switching bands and the shape of these bands correlates to the Q-factor of the corresponding eigenmodes and the modal responses to geometric variation.
This suggests the possibility of new optical-driven device design pathways which operate through a combination of eigenmode engineering and incident frequency tuning.

% \subsubsection{}
\section{Conclusion}

This study investigates lateral optical forces (LOFs) generated by periodic arrays of isosceles triangular motifs, where force emergence stems purely from geometric asymmetry.
Using Bayesian optimization, we identified three groups of structures exhibiting distinct optical force behaviors.
Notably, some geometrically dissimilar structures produce nearly identical optical forces response, while others show dramatic force variations, and direction reversals, from minimal geometric modifications.

To unravel this counterintuitive behavior, we mapped comprehensive landscapes of optical forces and diffraction efficiencies across the geometric parameter space.
These maps reveal two fundamentally different regions:
stable zones where forces remain robust against geometric variations, and switching bands where forces undergo abrupt transitions.

Through detailed spectral analysis of representative structures, we observed characteristic dip-and-peak lineshapes in the optical force spectra that strongly suggest Fano resonance behavior.
This recurring behavior points to a universal connection between switching bands and Fano-like responses, opening new possibilities for optical-driven device design.

The physical origin of this phenomenon was confirmed through eigenfrequency analysis of structures along a parameter-space loop.
The perfect correspondence between eigenmode excitations and switching band locations, along with the quantitative relationship between eigenmode Q-factors and transition rates, provides definitive evidence that these effects arise from Fano interference between discrete eigenmodes and continuum propagation states.

\begin{acknowledgments}
BG acknowledges the China Scholarship Council (CSC) and University of Bristol for a postgraduate scholarship (award number 202008060217),
and the Advanced Computing Research Centre (ACRC) at the University of Bristol for High Performance Computing platform BlueCrystal phase4 (BC4) support.
% HG acknowledges support through an Impact Acceleration Account co-funded by EPSRC and Bristol Nano Dynamics (grant number EP/R511663/1).
\end{acknowledgments}

\bibliography{ref}

\end{document}